\documentclass[useAMS,usenatbib,usedcolumn]{mn2e}

\usepackage{aas_macros}	
\usepackage{multirow}	
\usepackage{graphicx}
\usepackage{amsmath}	
\usepackage{amssymb}

\newcommand{\comment}[1]{} 
\newcommand{\ebv}{\ensuremath{E_{(B-V)}}}		
\newcommand{\kms}{\,km\,s$^{-1}$}
\newcommand{\pcc}{\,cm$^{-3}$}
\newcommand{\pcm}{\,cm$^{-2}$}
\newcommand{\hi}{H\,\textsc{i}}		
\newcommand{\htwo}{H$_2$}		
\newcommand{\ctwo}{C$_2$}		
\newcommand{\cai}{Ca\,\textsc{i}}	
\newcommand{\caii}{Ca\,\textsc{ii}}	
\newcommand{\caiii}{Ca\,\textsc{iii}}	
\newcommand{\ki}{K\,\textsc{i}}		
\newcommand{\nai}{Na\,\textsc{i}}	
\newcommand{\ci}{C\,\textsc{i}}		
\newcommand{\cii}{C\,\textsc{ii}}	
\newcommand{\kvel}{\ensuremath{\kappa}~Vel}

\usepackage{hyperref}
\hypersetup{colorlinks=true,linkcolor=red,citecolor=blue,filecolor=red,urlcolor=red,pdfauthor={Keith T. Smith et al.},pdftitle={Time-varying interstellar absorption towards κ Velorum}}

\title[Time-varying interstellar absorption towards $\kappa$~Velorum]{Small-scale structure in the interstellar medium: time-varying interstellar absorption towards $\boldsymbol{\kappa}$~Velorum}

\author[Keith T. Smith et al.]{
Keith T. Smith$^{1}$\thanks{E-mail:
\texttt{kts@ras.org.uk}; present address: Royal Astronomical Society, Burlington House, Piccadilly, London, W1J 0BQ, UK}, Stephen J. Fossey$^{2}$, Martin A. Cordiner$^{1,3}$, Peter J. Sarre$^{1}$,
\newauthor\ Arfon M. Smith$^{1,4}$, Tom A. Bell$^{5,6}$ and Serena Viti$^{2}$\\
$^1$School of Chemistry, The University of Nottingham, University Park, Nottingham, NG7 2RD, UK\\
$^2$Department of Physics and Astronomy, University College London, Gower Street, London, WC1E 6BT, UK\\
$^3$Astrochemistry Laboratory and The Goddard Center for Astrobiology, Mailstop 691, NASA Goddard Space Flight Center,\\ 8800 Greenbelt Road, Greenbelt, MD 20770, USA\\
$^4$Department of Physics, University of Oxford, Oxford, OX1 3RH, UK\\
$^5$Department of Astronomy, California Institute of Technology, Pasadena, CA 91125, USA\\
$^6$Centro de Astrobiolog\'ia (CSIC-INTA), 28850 Madrid, Spain
}

\begin{document}

\date{Accepted 2012 October 30.  Received 2012 October 27; in original form 2011
October 3}

\pagerange{\pageref{firstpage}--\pageref{lastpage}} \pubyear{2012}

\maketitle

\label{firstpage}

\begin{abstract}
Ultra-high spectral resolution observations of time-varying interstellar absorption towards $\kappa$~Vel are reported, using the Ultra-High Resolution Facility on the Anglo-Australian Telescope. Detections of interstellar \cai, \caii, \ki, \nai\ and CH are obtained, whilst an upper limit on the column density is reported for C$_2$. The results show continued increases in column densities of \ki\ and \cai\ since observations $\sim$\,4\,yr earlier, as the transverse motion of the star carried it $\sim$\,10\,AU perpendicular to the line of sight. Line profile models are fitted to the spectra and two main narrow components (A \& B) are identified for all species except CH. The column density $N(\text{\ki})$ is found to have increased by $82^{+10}_{-9}\%$ between 1994 and 2006, whilst $N(\text{\cai})$ is found to have increased by $32\pm5\%$ over the shorter period of 2002--2006. The line widths are used to constrain the kinetic temperature to $T_{k,\text{A}}<671^{+18}_{-17}$\,K and $T_{k,\text{B}}<114^{+15}_{-14}$\,K. Electron densities are determined from the \cai/\caii\ ratio, which in turn place lower limits on the total number density of $n_\text{A}\ga7\times10^3$\pcc\ and $n_\text{B}\ga2\times10^4$\pcc. Calcium depletions are estimated from the \cai/\ki\ ratio. Comparison with the chemical models of \citet{Bell2005} confirms the high number density, with $n=5\times10^4$\pcc\ for the best-fitting model. The first measurements of diffuse interstellar bands (DIBs) towards this star are made at two epochs, but only an upper limit of $\la40\%$ is placed on their variation over $\sim$\,9~years. The DIBs are unusually weak for the measured \ebv\ and appear to exhibit similar behaviour to that seen in Orion. The ratio of equivalent widths of the $\lambda5780$ to $\lambda5797$ DIBs is amongst the highest known, which may indicate that the carrier of $\lambda5797$ is more sensitive to UV radiation than to local density.
\end{abstract}


\section{Introduction}
Over the past two decades an increasing body of observational evidence has indicated that the diffuse interstellar medium (ISM) is highly structured on scales below one parsec, contrary to earlier theoretical expectations \citep{McKee1977} and observations which had suggested an apparent minimum cloud size \citep[e.g.][]{Crovisier1985}. Evidence for small-scale structure (SSS) has come from observations of interstellar absorption at radio, optical and ultraviolet wavelengths along closely separated lines-of-sight. Such sight-lines have been provided by spatially-resolved observations of extended extragalactic radio sources \citep{Dieter1976,Diamond1989,Lazio2009}, binary stars \citep{Meyer1990,Watson1996,Cordiner2006Fara} and star clusters \citep{Andrews2001,vanLoon2009}. Alternatively, a single background source can be observed repeatedly over several years, with each epoch probing a slightly different line-of-sight due to the transverse proper motion of the source and\slash{}or the intervening interstellar material. This technique has been applied to radio observations towards pulsars \citep{Frail1994,Stanimirovic2003,Stanimirovic2010} and optical observations towards nearby early-type stars (e.g. \citealt{Blades1997,Price2000,Danks2001,Welty2007}, see also the review by \citealt{Crawford2003}). These observations indicate that large differences in column density can exist between lines-of-sight with transverse separations from as few as tens to thousands of AU.

\citet{Heiles1997} considered possible theoretical scenarios to explain these variations without grossly violating pressure equilibrium with the surrounding ISM. His preferred model consisted of a population of cold and relatively dense structures in the form of sheets or filaments; the observed variations could then be produced if a structure was present and closely aligned to the line-of-sight. The model required very low temperature gas ($\sim15$\,K) to maintain pressure equilibrium. However, evidence for such a low temperature has not been found; instead more recent observations have indicated that a small but significant fraction of diffuse gas is at high pressure \citep{Jenkins2001,Jenkins2011}. \citet{Heiles2007obs} have summarised the observational and theoretical constraints on models for SSS, and theoretical attention has now shifted towards solutions which do not require dynamical equilibrium \citep[e.g.][]{Heiles2007theory}. Such over-pressured clouds are expected to evaporate quickly, so a mechanism (e.g. magnetohydrodynamic waves, as discussed by \citealt{Hartquist2003}) must be invoked to replenish the population of small-scale structures.

The existence of small-scale relatively dense regions within the diffuse ISM has important implications for interstellar chemistry. Although the interstellar radiation field is expected to be essentially unattenuated within such structures, the high density can result in enhancement of the molecular content \citep{Bell2005}. This provides a potential method for obtaining new information on the physical conditions, and a possible solution to some of the challenges which exist in reproducing the abundances of molecules observed along lines-of-sight through the diffuse ISM \citep{Cecchi2009}. A comprehensive understanding of the physical conditions within the SSS is required, so that it may be fed into such chemical models.

This paper is organised as follows: in section~\ref{sec:background} we describe previous studies of SSS towards \kvel, whilst our observations are detailed in section~\ref{sec:obs}. The results are presented in section~\ref{sec:results}, along with a line profile analysis. Section~\ref{sec:discuss} contains a comparison with earlier observations and a discussion of the physical and chemical conditions towards \kvel. Conclusions are presented in section~\ref{sec:conclusion}.

\section{Background}
\label{sec:background}
$\kappa$~Velorum (HD~81188) is a bright $V=2.46$ single-lined spectroscopic binary star with an orbital amplitude of 46.5\kms\ and period of 117 days \citep{Pourbaix2004}; the spectral type of the primary is B2\,IV, whilst the nature of the secondary is unknown. The distance to the system obtained from its trigonometric parallax measured by the \emph{Hipparcos} satellite is $165\pm13$\,pc \citep{Perryman1997}\footnote{\citet{vanLeeuwen2007} presented an updated reduction of the \emph{Hipparcos} data with significantly improved astrometry for most stars. However, in the case of \kvel\ the new reduction resulted in a `stochastic' solution as opposed to the fully converged solution found in 1997. The distances and proper motions calculated from the two reductions are consistent to within their mutual error bars. For this particular star we have therefore chosen to adopt the astrometry from \citet{Perryman1997}.}; the reddening is $\ebv=0.10$ \citep{Cha2000}. Early high-resolution observations of the ISM towards this star taken in 1989 and 1994 have been presented by \citet{Crawford1991} and \citet{Dunkin1999} respectively.

Small-scale structure towards \kvel\ was first noted by \citet{Crawford2000}, who serendipitously discovered that the equivalent width of the interstellar \ki\ line had increased by $\sim40$ per cent between 1994 and 2000, with a corresponding but smaller increase in absorption by \nai. The enhancement was found to be in a single narrow ($b=0.24$\kms) absorption component located at a velocity\footnote{All velocities quoted in this paper are heliocentric. To convert to LSR velocities towards \kvel\ subtract 11.43\kms.} $v_{\sun}=+8.5$\kms. The observations were interpreted as indicating that as the proper motion of \kvel\ carried it $\sim15$\,AU tangentially between 1994 and 2000 the line-of sight gradually intercepted a filament of cold and/or dense material aligned along the line-of-sight, of the type proposed by \citet{Heiles1997}.

Follow-up observations by \citet{Crawford2002} detected \cai\ and CH absorption at the same velocity, and found a continued increase in the \ki\ column density. From a consideration of the \cai/\caii\ ratio, CH abundance and stellar proper motion, \citet{Crawford2002} concluded that the absorbing material lay in a cool ($T$\,$\sim$\,$100$\,K), dense ($n_\text{H}$\,$\ga$\,$10^3$\pcc) cloud with approximate dimensions of $100\text{--}1000$\,AU in depth (along the line of sight) and 15\,AU in the transverse direction.

Consideration of this cloud prompted \citet{Bell2005} to develop a chemical model of such a filament. The column densities of several atomic and molecular species were calculated for a cloud with physical extent similar to that inferred by \citet{Crawford2002} as a function of time, radiation field and density. The models which most closely reproduced the observations towards \kvel\ were for a young ($\sim50$\,yr), transient (lifetime $<100$\,yr), and high density ($n_\text{H}\ga10^4$\pcc) filament immersed in a typical ambient interstellar radiation field. The models predicted that detectable column densities of additional diatomic molecules could be present, including C$_2$ and OH.

The line of sight towards \kvel\ offers an excellent opportunity to explore the on-going chemical and dynamical evolution of small-scale structure. We therefore set out to re-observe the chemical species studied by \citet{Crawford2002}, and to conduct a search for C$_2$, in order to further constrain the physical and chemical conditions and test the chemical model of \citet{Bell2005}. In particular, detection of the homonuclear molecule C$_2$ could be used to measure the local kinetic temperature through the populations of each rotational $J$-state (and radiation temperature and density, if sufficient high-$J$ lines were also detected, cf. \citealt{vanDishoeck1982}). The resolved intrinsic profiles of the narrow interstellar lines could then be used to infer the turbulent contribution to the line widths \citep{Crawford1997}, which could be particularly important for understanding the origin and evolution of the small-scale structure.

\section{Observations}
\label{sec:obs}

\begin{table}
	\centering
	\caption[Summary of the UHRF observations]{Summary of the UHRF observations towards \kvel. Exp time is the total exposure time, S/N is the resulting per-pixel continuum signal-to-noise ratio and $W$ is the equivalent width integrated over the entire velocity profile. Errors are $1\sigma$. The \ctwo\ upper limit is $3\sigma$ on each of the five lines which fall in the observed wavelength region.}
	\label{tab:uhrfobs}
	\begin{tabular}{llD{.}{.}{2}lc}
		\hline
			& UT date 	& \multicolumn{1}{c}{Exp time}	& S/N 	& $W$ \\
			&			& \multicolumn{1}{c}{(hr)} 		& 		& (m\AA) \\
		\hline
		\cai & 2006 March 10 & 4.0 & 380 & $1.13\pm0.02$\\
		\caii & 2006 March 11 & 0.5 & 160 & $23.24\pm0.26$~~\\
		\ki & 2006 March 09,10 & 2.75 & 210& ~$7.49\pm0.34$~~\\
		\nai & 2006 March 11 & 0.67 & 190 & $60.18\pm0.14$~~\\ 
		CH & 2006 March 09,11 & 9.0 & 600 & $0.25\pm0.03$\\
		\ctwo & 2006 March 12-15 & 15.0 & 800 & $<0.11$\\
		\hline
	\end{tabular}
\end{table}

Observations of interstellar absorption towards \kvel\ were made on the nights of 2006~March~9--15 using the Ultra-High Resolution Facility (UHRF, \citealt{Diego1995}) located at the coud\'e focus of the 3.9~m Anglo-Australian Telescope (AAT). The UHRF is an \'echelle spectrograph which is capable of ultra-high resolution but can only record simultaneous data over a very narrow wavelength region, so a separate setup is required for each species. A summary of the observations is presented in table~\ref{tab:uhrfobs}. New observations of the absorption lines due to \cai, \caii\ (K line), \ki\ (7698\,\AA\ line) and \nai\ (D$_1$ line) were obtained, along with rotational lines of the A$^{2}\Delta \leftarrow$~X$^{2}\Pi$ (0,0) band of the CH molecule in the R$_2$(1) $\Lambda$-doublet (comprising the R$_{fe}$(1/2) and R$_{ff}$(1/2) lines). A search was performed for the \ctwo\ molecule in several low-$J$ lines (R(0), Q(2), Q(4), Q(6) and P(2)) of the A$^{1}\Pi_u \leftarrow$~X$^{1}\Sigma^{+}_{g}$ Phillips (2,0) band, which were covered in a single instrumental set-up.

The detector used for the atomic lines and CH was the blue-optimised EEV2 CCD ($2048 \times 4096$ 13.5-\micron\ pixels). The EEV2 was operated in normal readout mode, which provides an RMS readout noise of 3.2\,e$^-$/pixel and an unbinned readout time of 180 seconds. Because the image slicer (see below) projects the spectrum onto $\sim 1400$ unbinned pixels, the EEV2 was binned by a factor of eight perpendicular to the dispersion to reduce the readout noise and time. The EEV2 CCD suffers from severe fringing at wavelengths above $\sim6500$\,\AA\ which particularly affected the \ki\ observations, necessitating careful flat-fielding. For the C$_2$ observations the red-optimised MITLL3 CCD ($2048 \times 4096$ 15-\micron\ pixels) was used due to its higher quantum efficiency and lower fringing at long wavelengths. As the MITLL3 detector could not be binned it was operated in slow readout mode, which provides a readout noise of 1.8\,e$^-$/pixel and a readout time of 300\,s.

The UHRF was operated in `R\,=\,1E6' mode, the highest spectral resolution available, which is vital to study the profiles of narrow interstellar lines. Wavelength calibration used a Th-Ar arc lamp, whilst flat fields were obtained with an internal quartz-halogen calibration lamp. In this resolution mode the instrument resolves the lines from the arc lamp, so a stabilised He-Ne laser was used to measure the instrumental resolution. The resolving power ($R\equiv\lambda/\Delta\lambda$) was found to be $R\simeq845,000$ for the EEV2 and $R\simeq940,000$ for the MITLL3, corresponding to velocity resolutions of 0.355 and 0.319\kms\ respectively.

Individual exposures were limited to 30\,min to prevent degradation of the instrumental resolution by the changing heliocentric velocity correction due to the rotation of the Earth. The detector readout times therefore constituted a significant overhead, and detector readout noise made a small but significant contribution to the overall noise level.

The UHRF incorporates an image slicer which reformats a $1.5\arcsec\times1.5$\arcsec\ entrance aperture to a 0.06\arcsec\ slit to reduce slit losses, but itself has a peak throughput of only 15\%. Observing conditions were clear on the nights of 9 and 15 March, but those of 10--14 March were frequently interrupted by cloud and fog. The seeing during the observations varied in the range 1--3\arcsec, mostly around 1.5\arcsec.

Data reduction was performed using the \texttt{noao.onedspec} routines in \textsc{iraf}\footnote{\textsc{iraf} is distributed by the National Optical Astronomy Observatories, which are operated by the Association of Universities for Research in Astronomy, Inc., under cooperative agreement with the National Science Foundation.}. Pixel-to-pixel sensitivity variations and fringing were removed by division by a normalised flat-field. The scattered light background was measured from the inter-order regions and subtracted. The \nai, \cai, \ki, CH and C$_2$ observations were corrected for telluric absorption by division by a well-exposed spectrum of a bright, fast-rotating, early-type standard star ($\alpha$~Vir and/or $\alpha$~Eri); the \caii\ observations were found to be free of telluric contamination. Individual exposures were converted to a heliocentric reference frame before being co-added and converted to a velocity scale.

Additional observations of diffuse interstellar bands \citep[DIBs -- see the review by][]{Sarre2006} were obtained using other spectrographs at two additional epochs. The earlier observations were obtained during January 1995 as part of a survey of DIBs towards Southern stars conducted by one of us (SJF). The instrument used was the coud\'e \'echelle spectrograph on the 74-inch telescope at the Mount Stromlo Observatory. The resolving power was $R\simeq60,000$ and the signal-to-noise was $\sim550$. Later observations were obtained during June 2004, as part of a larger study of DIBs as tracers of SSS (\citealt{Cordiner2006Fara}, Cordiner et al. in prep). The instrument used was the UCLES spectrograph on the AAT. The spectral resolution was $R\simeq58,000$ and the signal-to-noise was $\sim2,300$. Full details of the observations and data reduction have been given by \citet{Cordiner2006Thesis} and \citet{Cordiner2006Fara}. The 1995 and 2004 data were corrected for telluric absorption using composite spectra obtained by co-adding spectra of the same two stars ($\alpha$~Vir and $\alpha$~Eri), observed at both epochs. These telluric standards were obtained at similar airmasses to \kvel\ and required negligible intensity scaling, so any residual stellar features should be the same at both epochs.

\section{Results \& line profile analysis}
\label{sec:results}


\begin{figure*}
	\centering
		\includegraphics[width=\textwidth]{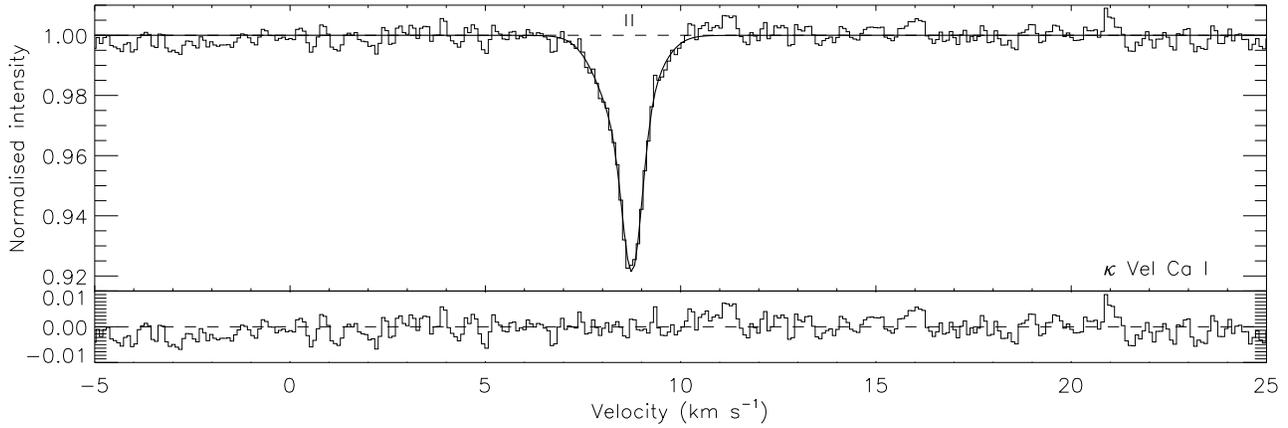}
	\caption{Interstellar \cai\ absorption observed towards \kvel. In the upper panel, the thin histogram is the observed spectrum and the thick line is the fitted model profile. The lower panel shows the residuals. The dashed line indicates the continuum level and tick marks indicate the positions of each absorption component listed in table \ref{tab:uhrfmodels}. The velocity scale is heliocentric.}
	\label{fig:cai}
\end{figure*}

\begin{figure*}
	\centering
		\includegraphics[width=\textwidth]{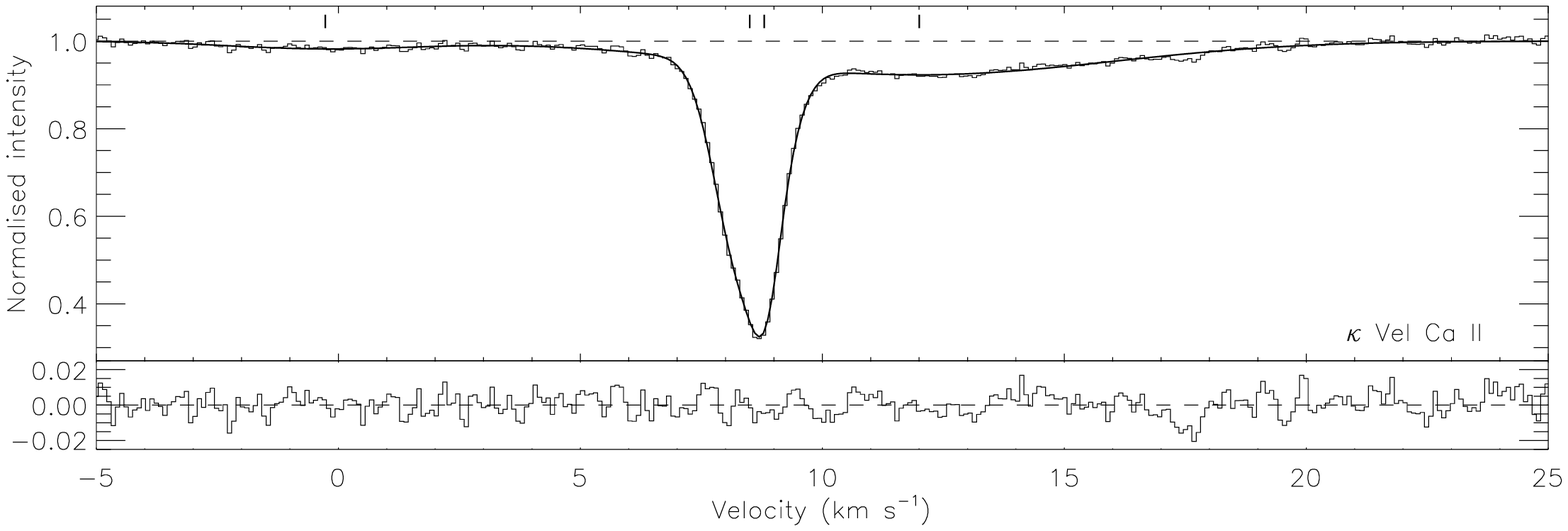}
	\caption{As in figure~\ref{fig:cai}, but for interstellar \caii~K absorption.}
	\label{fig:caii}
\end{figure*}

\begin{figure*}
	\centering
		\includegraphics[width=\textwidth]{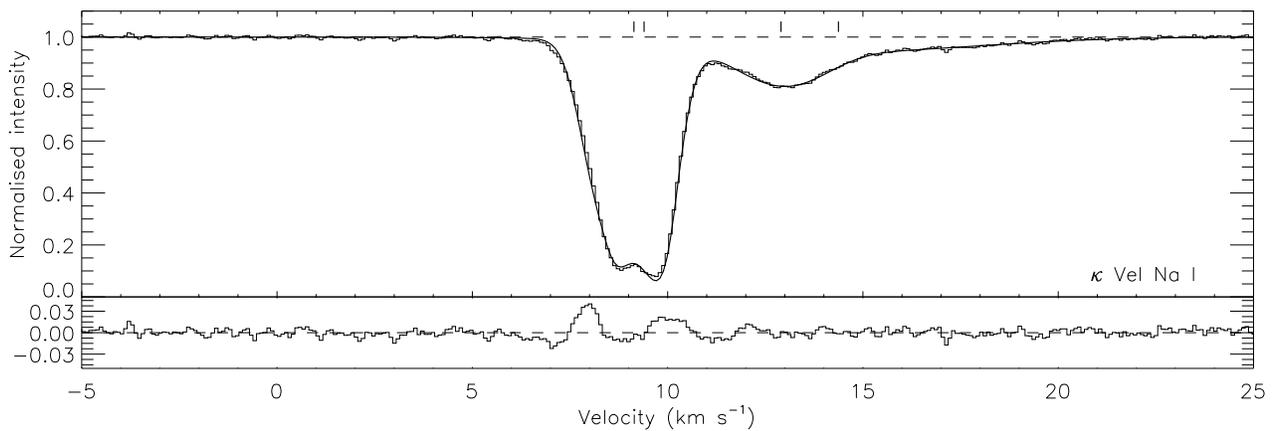}
	\caption{As in figure~\ref{fig:cai}, but for interstellar \nai~D$_1$ absorption. The model does not completely fit the core of the line, see section~\ref{sec:nai}.}
	\label{fig:nai}
\end{figure*}

\begin{figure*}
	\centering
		\includegraphics[width=\textwidth]{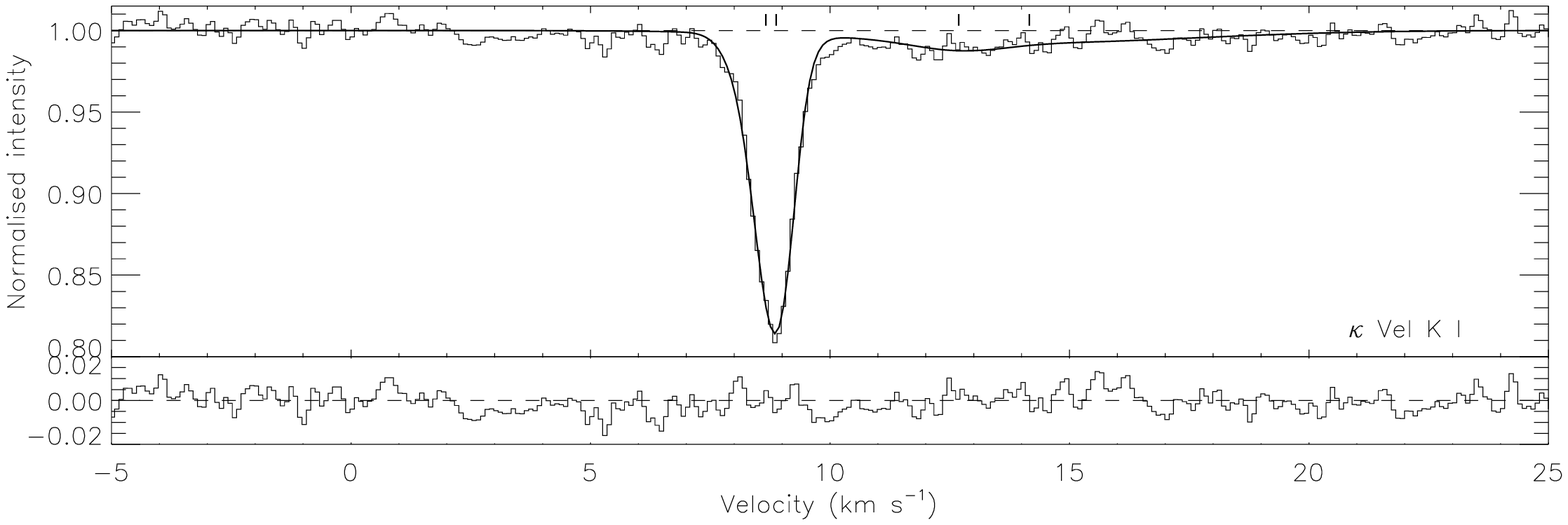}
	\caption{As in figure~\ref{fig:cai}, but for interstellar \ki\ absorption.}
	\label{fig:ki}
\end{figure*}

\begin{figure*}
	\centering
		\includegraphics[width=\textwidth]{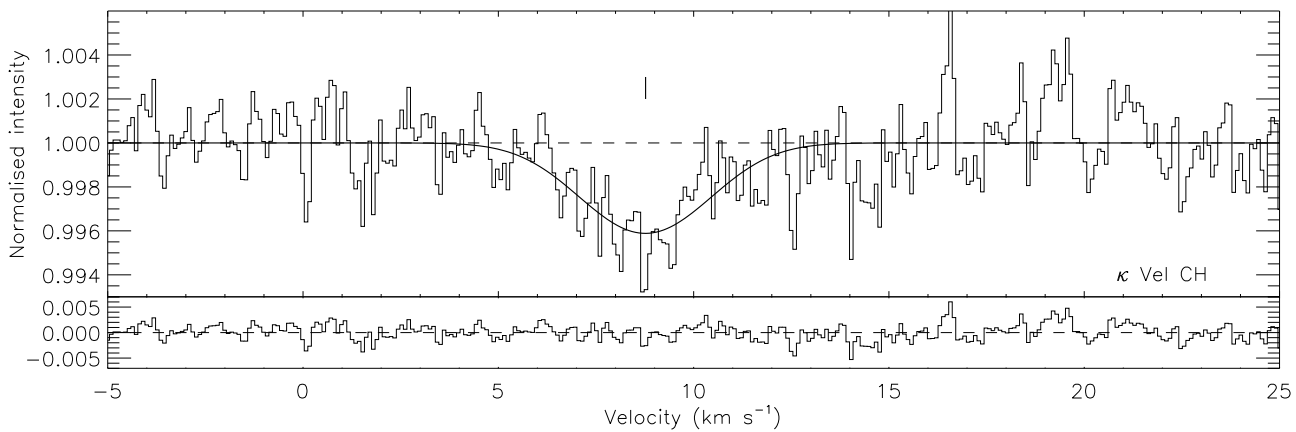}
	\caption{As in figure~\ref{fig:cai}, but for interstellar CH absorption.}
	\label{fig:CH}
\end{figure*}

High signal-to-noise (S/N) was achieved in the UHRF observations of each line, improving upon the earlier observations of \citet{Crawford2002} by factors of 1.5--2 in S/N. All of the atomic absorption lines were detected at high significance. Absorption by the CH molecule was also found but none of the \ctwo\ lines were detected, for which we determined upper limits. The UHRF spectra are shown in figures \ref{fig:cai}--\ref{fig:CH}; the continuum signal-to-noise ratios and measured total equivalent widths are given in table \ref{tab:uhrfobs}. Preliminary results from these observations were presented by \citet{Smith2010}; they are superseded by the results presented here.

The line profiles of the detected species were modelled using the \textsc{vapid} interstellar line modelling program \citep{Howarth2002}, which computes the optimum column density $N$, velocity $v$ and Doppler broadening parameter\footnote{The broadening parameter $b$ is related to the FWHM of the corresponding Gaussian distribution by $b=\frac{\text{FWHM}}{2\sqrt{\ln{2}}}$} $b$ for a given set of absorption components using a non-linear $\chi^2$ minimisation algorithm. Individual absorption components have a Voigt profile formed by the convolution of the Lorentzian natural line shape, the instrumental point-spread-function (taken to be Gaussian with widths given by the velocity resolution found in  section~\ref{sec:obs}) and a Gaussian with broadening parameter $b$.

\begin{table}
	\centering
	\caption[Atomic and molecular line data]{Atomic and molecular line data adopted in the line profile analysis. Wavelengths and oscillator strengths $f$ are taken from \citet{Morton2003} for the atomic species, \citet{Black1988} for CH and \citet{Sonnentrucker2007} for \ctwo. The \nai\ and \ki\ lines have hyperfine structure, whilst C$_2$ and CH each have several rotational lines. Oscillator strengths have been apportioned between hyperfine components in a 5:3 ratio according to the prescriptions of \citet{Welty1994}. For \ctwo, the line ratios given by \citet{Sonnentrucker2007} have been adopted.}
	\label{tab:uhrfatomicdata}
	\begin{tabular}{llcc}
		\hline
						& 				& Wavelength	& $f$	\\
						&					& (\AA)				&			\\
		\hline
		\cai\ & & 4226.728 & $1.77$ \\
		\\
		\caii\ & K & 3933.6614 & $6.267\times10^{-1}$ \\
		\\
		\ki\ & & 7698.9586 & $1.248\times10^{-1}$ \\
		 & & 7698.9681 & $2.079\times10^{-1}$ \\
		\\
		\nai\ & D$_1$ & 5895.9109 & $1.200\times10^{-1}$\\
		 & & 5895.9322 & $2.001\times10^{-1}$\\
		\\
		CH & R$_{ff}$(1/2)& 4300.3235 & $2.53\times10^{-3}$\\
		 & R$_{fe}$(1/2)& 4300.3030 & $2.53\times10^{-3}$\\
		\\
		\ctwo\ & R(0) & 8757.686 & $1.40\times10^{-3}$ \\
		 & Q(2) & 8761.194 & $7.0\times10^{-4}$ \\
		 & Q(4) & 8763.751 & $7.0\times10^{-4}$ \\
		 & Q(6) & 8767.759 & $7.0\times10^{-4}$ \\
		 & P(2) & 8766.031 & $1.4\times10^{-4}$ \\
		\hline
	\end{tabular}
\end{table}

The wavelengths and oscillator strengths adopted in the analysis are given in table~\ref{tab:uhrfatomicdata}, and were taken from \citet{Morton2003} for the atomic species, \citet{Black1988} for CH and \citet{Sonnentrucker2007} for C$_2$. The CH rotational lines form a $\Lambda$-doublet split by 1.43\kms; the velocities output by \textsc{vapid} are with respect to the bluest transition, and were converted to the weighted mean of the two lines by subtraction of 0.72\kms. The two levels which contribute to the doublet were assumed to have equal populations, because the energy splitting between them is negligibly small compared to the thermal energy \citep[cf.][]{Crane1995}; the oscillator strengths used in \textsc{vapid} were apportioned accordingly. The \ki\ and \nai~D$_1$ transitions are split by nuclear hyperfine structure, resulting in two lines 0.370 and 1.083\kms\ apart respectively. The relative strengths of the transitions are equal to $2F+1$, which for both observed \nai\ and \ki\ transitions gives a ratio of 5:3 with the stronger transition being the redder of the pair (see \citealt{Welty1994} or \citealt{Morton2003} for derivations). The \textsc{vapid} output velocities were converted to the weighted mean velocity of the transitions by subtraction of 0.231\kms\ and 0.677\kms\ for \ki\ and \nai\ respectively. For each \ctwo\ $J$-state, the oscillator strength was apportioned between the P, Q and R branches according to the relative line strengths given by \citet{Sonnentrucker2007}. The results of the fitting are presented in table~\ref{tab:uhrfmodels}.

%

\begin{table*}
	\centering
	\caption[Results of the line profile analysis]{Results of the line profile analysis of the UHRF observations of interstellar absorption towards \kvel. The heliocentric velocity $v_{\sun}$, Doppler broadening parameter $b$ and logarithm of the column density $N$ (in \pcm) are given for each model component. The total convolved line profiles are plotted in figures \ref{fig:cai}--\ref{fig:CH}. $T_k^{\rm{ul}}$ is the upper limit on the kinetic temperature calculated from equation \ref{eq:b} assuming no turbulent contribution to the $b$-value (see section~\ref{sec:temp}). Heliocentric velocities are quoted with respect to the weighted means of the transitions for each species. Statistical uncertainties are quoted as $1\sigma$.}
	\label{tab:uhrfmodels}
	\begin{tabular}{lcccc}
		\hline
		& $v_{\sun}$ & $b$ & $\log{N}$ & $T_k^{\rm{ul}}$ \\
		& (\kms) & (\kms) &  & (K)\\
		\hline
		\smallskip
		\cai &$^\text{A}8.604\pm0.033$ & $0.851\pm0.049$ & $9.412\pm0.032$ & $1760^{+210}_{-200}$\\
		 & $^\text{B}8.762\pm0.011$ & $0.268\pm0.025$ & $9.198\pm0.051$ &  $174^{+34}_{-31}$\\
		 \\
		\caii & $-0.27\pm0.18$ & $2.81\pm0.21$ & $10.123\pm0.033$ & -\\
		 & $^\text{A}8.500\pm0.005$ & $0.755\pm0.006$ & $11.240\pm0.005$ & $1381^{+22}_{-22}$\\
		 & $^\text{B}8.802\pm0.007$ & $0.283\pm0.012$ & $10.521\pm0.024$ & $194^{+17}_{-16}$\\
		 & $12.003\pm0.068$ & $5.516\pm0.086$ & $11.078\pm0.006$ & -\\
		 \\
		\ki & $^\text{A}8.200\pm0.017$ & $0.609\pm0.036$ & $10.141\pm0.014$ & $880^{+110}_{-100}$\\ 
		 & $^\text{B}8.414\pm0.008$ & $0.346\pm0.017$ & $10.234\pm0.009$ & $283^{+28}_{-27}$\\ 
		 & $12.225^f$ & $1.31^f$ & $9.30\pm0.12$ & -\\
		 & $13.70^f$ & $4.86^f$ & $9.979\pm0.050$ & -\\
		 \\
		\nai & $^\text{A}8.460\pm0.007$ & $0.694\pm0.009$ & $11.887\pm0.009$ & $671^{+18}_{-17}$\\ 
		 & $^\text{B}8.720\pm0.008$ & $0.286\pm0.018$ & $11.327\pm0.032$ & $114^{+15}_{-14}$\\ 
		 & $12.225\pm0.039$ & $1.31\pm0.11$ & $10.933\pm0.053$ & -\\ 
		 & $13.70\pm0.71$ & $4.86\pm0.55$ & $11.007\pm0.092$ & -\\ 
		 \\
		CH & ~$8.77\pm0.23$ & $2.06\pm0.56$ & $11.470\pm0.051$ & -\\
		\hline
		\multicolumn{5}{l}{$^\text{A}$ component A, $^\text{B}$ component B,}\\
		\multicolumn{5}{l}{$^f$ values held fixed during fitting (see text)}
	\end{tabular}
\end{table*}

\subsection{Uncertainties}
\label{sec:uncertainties}
The results of the line profile analysis are affected by a number of sources of uncertainty, both statistical and systematic. Rigorous estimates of the statistical uncertainty were obtained by generating 1,000 Monte-Carlo noise simulations for each line, in which Gaussian noise corresponding to the measured signal-to-noise ratio was added to the data and the fits repeated. The lower signal-to-noise present in the cores of the lines due to the fewer photons detected was taken into account during the Monte-Carlo process, under the assumption of Poisson statistics. Statistical uncertainties are given in table~\ref{tab:uhrfmodels}.

In addition to these statistical uncertainties, there are a number of systematic uncertainties, primarily affecting the measured velocities. The small wavelength coverage of the UHRF (4--9\,\AA) results in only a low number of arc lines for wavelength calibration which limits the absolute precision, particularly in the case of \ctwo\ for which only four arc lines were available. The magnitude of this effect can be estimated from the RMS of the wavelength fit (which used a cubic polynomial, quadratic for \ctwo), which was always less than 0.01\kms. Uncertainties in the atomic data listed in table~\ref{tab:uhrfatomicdata} may affect the velocities e.g. the wavelength of the \caii~K line is known to be uncertain by about 0.02\kms\ \citep{Morton2003}. Velocity shifts \emph{between} observations of a few tenths of a \kms\ have been seen in previous UHRF data \citep[e.g.][]{Crawford2000} and are thought to result from slight differences between the light paths of the telescope and arc lamp beams. The slight velocity offsets between the velocity components found in this work and those found by \citet{Crawford2000} are of similar magnitude. The quoted velocities are therefore systematically uncertain at around this level. Nevertheless, the multiple components for each species should be affected in an identical fashion by this systematic uncertainty, so the \emph{relative} velocities between components of a single species are unaffected, and so we do not incorporate this systematic uncertainty into table~\ref{tab:uhrfmodels}.

Another potential source of systematic uncertainty is the definition of the continuum level. With the exception of \caii\ (see section~\ref{sec:caii}) there are no stellar counterparts to the interstellar lines. Around a thousand pixels of clear continuum were present on both sides of every observed line. This greatly eases determination of the continuum level, which was fitted with a low-order polynomial (quadratic or cubic). The uncertainty in the continuum level is therefore neglible compared to the statistical uncertainties.

Finally, we consider the systematic uncertainty introduced by the determination of the instrumental point-spread function (PSF). Inspection of the He-Ne laser and Th-Ar arc lamp spectra shows no evidence for a non-Gaussian PSF. The measurement of the width of the PSF (and hence instrumental resolution) carries an uncertainty of $\sim0.001$\kms, which translates to a systematic uncertainty of $\sim0.0006$\kms\ on the measured $b$ values. This uncertainty is negligibly small compared to the statistical uncertainties on $b$, and affects all model components equally, so it has not been included in the uncertainties quoted in table~\ref{tab:uhrfmodels}.

\subsection{Atomic species}
\label{sec:atomic_models}
Previous observers have modelled the main absorption around 9\kms\ with either one or two separate components. \citet{Dunkin1999} and \citet{Crawford2000} found that their \ki\ and \nai\ data required two components, whilst \citet{Crawford2002} modelled only a single component in \cai\ and CH (but noted that these line profiles would also be consistent with the presence of two components, as found for the stronger lines). Whilst modelling the atomic species, it was found that a two-component model fitted to the narrower $\sim9$\kms\ feature was significantly better than a one-component model for all atomic species. This was confirmed through examination of the fit residuals and via a statistical F-test, which found that the two component model was statistically better (in the sense of improvement in $\chi^2$) at least the 95\% confidence level. The similarities in the velocities and $b$ values obtained for each species, which were derived independently (except for \ki, see below), reinforces the interpretation that a two component model is justified by the new high signal-to-noise data.

For each atomic species, the main absorption around 9\kms\ contains a narrow component at $v\sim8.5$\kms\ with $b\sim0.7$\kms\ (hereafter component A) and a slightly higher velocity even narrower component at $v\sim8.7$\kms\ with $b\sim0.3$\kms\ (component B). These components are consistent with the previously reported two-component $b$ values of \citet{Crawford2000}, though at slightly shifted velocity. All of the atomic species except \cai\ also include a broader absorption around $\sim12$\kms; this absorption was seen in \caii\ and \ki\ by \citet{Crawford2002}, but only included in his \caii\ model. The methodology adopted in modelling each species is described in more detail below.

\subsubsection{\cai}
There is no evidence for broad components in the \cai\ absorption, so the model consists of the narrow components A \& B only (see figure~\ref{fig:cai}). Although the absorption only reaches a depth of 8\% of the continuum, the high signal-to-noise data confirm the need for two narrow components.

\subsubsection{\caii}
\label{sec:caii}
Calcium~K is the only observed line which has a photospheric counterpart. The rotational velocity of \kvel\ is $v_{\text{rot}}\sin{}i=55$\kms\ \citep{Glebocki2000}, so the stellar line is much broader than the interstellar features. Fortuitously, the orbital motion of the spectroscopic binary during our observations resulted in displacement of the stellar line to a velocity well clear of the interstellar absorption. In contrast to previous observations, the stellar line does not overlap the interstellar features in our data, which makes the continuum definition straightforward. The narrow absorption around 9\kms\ clearly requires two components in the model (see figure~\ref{fig:caii}). The broad higher-velocity absorption has been modelled with a single $b=5.516\pm0.086$\kms\ component (\citealt{Crawford2002} also modelled this as a single component). There is no evidence for two broad components as found for \nai\ (see section~\ref{sec:nai} below). A new \caii\ absorption component at about 0\kms\ is also found, which was regarded as part of the stellar continuum by \citet{Crawford2002}.

\subsubsection{\nai}
\label{sec:nai}
For this species it was found that two broad higher-velocity cloud components were required in addition to the two narrow components A and B (see figure~\ref{fig:nai}). These components were also found by \citet{Crawford2000}, though the parameters of the broad components differ in detail, typically at the 2-$\sigma$ level (but less so when compared to the 1994 observations). Our 12\kms\ component is slightly narrower and weaker, and our 14\kms\ component stronger and broader, than \citeauthor{Crawford2000}'s corresponding components; however the total column density of the two broad components has in fact changed very little between 1994, 2000 and 2006. Given their marginal significance, we therefore consider that the small differences between the modelled broad components might simply be due to the covariance between the fitted parameters, given the difficulty in defining accurately the velocities of such broad overlapping components. If the $\sim14$\kms\ component is instead fixed to the velocity found by \citet{Crawford2000}, all other fitted parameters become quite consistent with the other epochs, for both broad components. There is therefore no evidence that these broad components are varying between epochs.

There remains a slight discrepancy between the \nai\ model and data at the core of the narrow components, indicating that further substructure may be present. However, models with additional components did not improve the fit, because the (superposed) blueward hyperfine component within the line cannot be reproduced without the redward hyperfine component becoming too deep relative to the data. An investigation was undertaken to determine whether this behaviour could be explained by an error in the zero point of the spectrum, but adjusting the level of the zero point did not improve the fit.

\subsubsection{\ki}
For potassium a broad high-velocity absorption is visible (figure~\ref{fig:ki}). This can be adequately modelled with a single component, but with a very high $b$ value ($\sim$\,7\kms). However, because this absorption is seen to consist of two separate components in \nai, which is expected to be similarly distributed to \ki\ \citep{Pan2005}, we have chosen to reproduce these two components in the \ki\ model. Although the two components do not result in a statistically significant improvement in the fit, the two-component model is more physically plausible, given the \nai\ model, and avoids depressing the continuum around the narrow components A and B. The broad components in the \ki\ model were fixed at the velocity and $b$-values found for \nai, but were allowed to vary in column density.

In an unconstrained fit, the narrow components (A and B) collapse to the same velocity, and the model becomes inconsistent with the velocity separation seen in the other atomic species. We therefore prefer a model for \ki\ in which the components reproduce the velocity separation seen in earlier observations and in the other species. We iterated towards an acceptable model as follows: first the correlation between the distribution of \ki\ and \nai\ which was assumed for the broad components was extended to the narrow components. The velocity separation between A and B was constrained to be consistent with both the other species and the two-component \ki\ model of \citet{Crawford2000}. When the column densities and $b$-values in the $\sim9$\kms\ component were fitted in this model, component A appears to be weaker ($\log N_\text{A} = 9.99$) than the 1994 value reported by \citet{Crawford2000}, with B somewhat stronger ($\log N_\text{B} = 10.32$). \citet{Crawford2000} found that their results were consistent with only the redward component (B) having varied between 1994 and 2000, so we also fixed the column density of component A to the 1994 value, to obtain the final model presented in table~\ref{tab:uhrfmodels} (the uncertainties on the parameters of component A were obtained by fixing all other components at their fitted values, so are likely underestimated). In all unconstrained fits, component B is even stronger than tabulated here. However, the total column density in both narrow components (A+B) is identical whichever approach is adopted, so that figure is robust even if the relative strengths remain uncertain.

\subsubsection{Comparison between atomic species}
Although the two main absorption components are broadly consistent between atomic species, their exact velocities, $b$-values and the velocity separation of the two components are not. These differences are most likely to be due to the systematic uncertainties discussed in section~\ref{sec:uncertainties}. Nevertheless, the \emph{relative} velocities between components of the same species are secure.

The high-velocity components seen in \nai, \ki\ and \caii\ have high $b$-values of up to 5\kms. There are three possible explanations for such high $b$-values:
\\a) High temperature gas. We reject this possibility because the temperatures required ($\gtrsim$\,30,000\,K) are inconsistent with the presence of substantial amounts of neutral \ki\ and \nai.
\\b) Very turbulent gas, with $v_t\gtrsim4$\kms. This would be an unusually high turbulent velocity, but cannot be excluded by the current data.
\\c) Multiple blended absorption components, which cannot be distinguished at the current signal-to-noise ratio. This seems the most likely explanation \citep[cf.][]{Dunkin1999}.

The differences in separation between the narrow components may indicate intrinsic unresolved structure within the absorbing material. Whilst the absorption has been modelled as two components, this does not necessarily imply that there are two physically separate clouds along the line of sight. In particular, the electron and number densities derived in section~\ref{sec:discuss} are anomalously high for \emph{both} components. There may exist a range of physical conditions within a single `cloud', which are approximated by the two component model. The velocity separation between components A and B are larger for \caii\ and \nai\ than for \cai\ (and \ki). This could potentially be explained if \caii\ and \nai\ are present throughout the cloud, whilst \cai\ is present only in the denser regions and \ki\ is intermediate between \nai\ and \cai\ (the distribution of these species in diffuse clouds has been discussed by \citealt{Pan2005}). If so, the physical parameters derived in section~\ref{sec:discuss} are indicative of the range of conditions throughout the cloud, but will exclude the most extreme regions. This may also explain the imperfect fit of the model to the core of the \nai\ line.

\subsection{CH}
\label{sec:ch}
Whilst it might have been expected that CH would have a similar distribution to the atomic species, perhaps with enhancement in the cooler (and presumably denser) component B, no evidence for this has been found. There is no statistical justification for more than one component in the current data. The CH line profile was modelled with a single, broad ($b=2.06$\kms) absorption component, albeit with a high uncertainty ($\sigma_b=0.56$\kms), as shown in figure~\ref{fig:CH}. The absorption profile has a central depth of less than 0.5\% of the continuum. Although the velocity of the CH absorption is closest to that of component B, the uncertainty and variations between species are large, so CH cannot be conclusively identified with any component seen in the atomic species. We consider the breadth of the CH line profile in more detail in section~\ref{sec:CH_sim}.

\subsection[C_2]{\ctwo}
\label{sec:c2_obs}
The continuum signal-to-noise ratio achieved in the region of the \ctwo\ lines was $\sim$\,800 but no lines were found. To improve the chances of a detection, the expected positions of the five rotational lines were stacked in velocity space, but this also failed to reveal any absorption. A $3\sigma$ upper limit of 0.11\,m\AA\ was determined on each of the five lines. The corresponding upper limit on the total \ctwo\ column density is model dependent; we return to this point in section~\ref{sec:c2_col}.

\subsection{Diffuse interstellar bands}

\begin{table}
\centering
\caption[]{Measured equivalent widths ($W$) of DIBs and atomic lines towards $\kappa$ Vel in January 1995 and June 2004. In the final column, `Y' indicates a definite or probable stellar blend contaminating the DIB, `--' indicates improbable or absent stellar blend and `?' is for indeterminate cases. Uncertainties and upper limits are $\sigma_c\Delta\lambda$.}
\label{tab:dibs}
\begin{tabular}{lrrc}
\hline
 & \multicolumn{2}{c}{$W$ (m\AA)} & \\
\cline{2-3}
 & Jan 1995 & June 2004 & Blend?\\
\hline
$\lambda$5780 & $10\pm4$ & $10\pm1$ & ?\\
$\lambda$5797 & $<1.5$ & $< 0.4$ & ?\\
$\lambda$5850 & $<1.2$ & $<0.4$ & --\\
$\lambda$6196 & $<0.6$ & $1.1\pm0.2$ & --\\
$\lambda$6203 & $<6$ & $3\pm2$ & --\\
$\lambda$6283 & * & $19\pm6$ & Y\\
$\lambda$6613 & $<1.8$ & $3\pm0.5$ & --\\
K~\textsc{i} 7698 & $\dagger$ & $5.5\pm0.2$ & -- \\
Na~\textsc{i} D$_1$ & $60.0\pm0.3$ & $62.2\pm0.2$ & -- \\
\hline
\multicolumn{4}{l}{* Severe telluric contamination}\\
\multicolumn{4}{l}{$\dagger$ Outside observed wavelength range}
\end{tabular}
\end{table}

\begin{figure}
	\centering
		\includegraphics[width=\columnwidth]{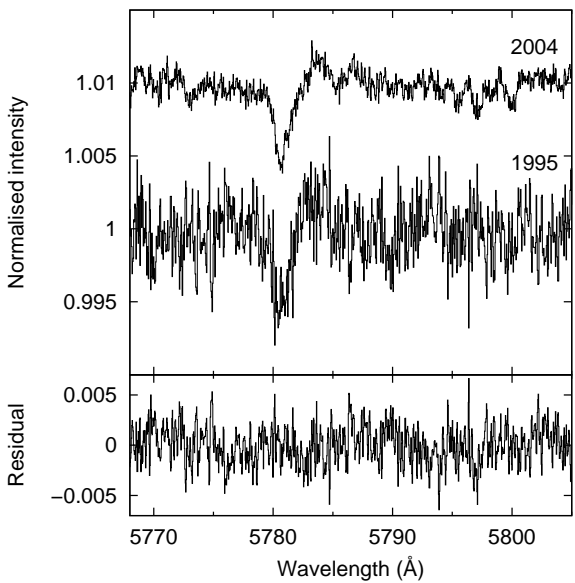}
	\caption{Spectra of the region around the $\lambda$5780 and $\lambda$5797 diffuse interstellar bands recorded towards \kvel. The upper trace is the observed spectrum in June 2004, the middle trace is the same region as observed in January 1995, and the lower trace shows residual intensities of the 2004 minus the 1995 data. The $\lambda$5797 DIB is not detected at either epoch, and there is no evidence for temporal variation of the $\lambda$5780 DIB.}
	\label{fig:5780}
\end{figure}

\begin{figure}
	\centering
		\includegraphics[width=\columnwidth]{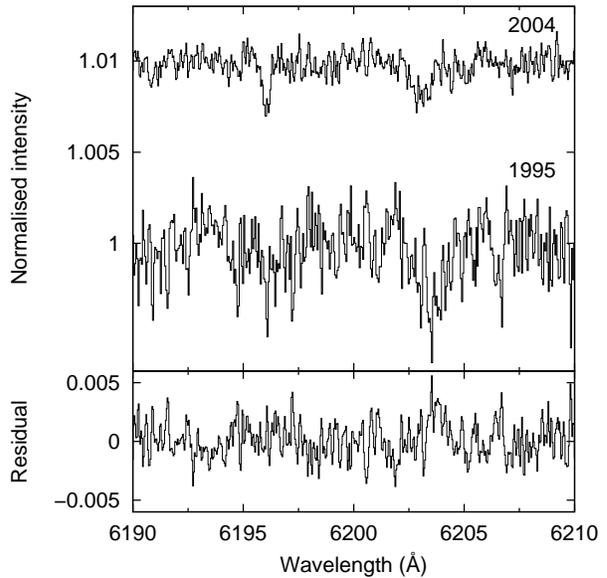}
	\caption{As in figure~\ref{fig:5780}, but for the $\lambda$6196 and $\lambda$6203 DIBs. Both DIBs are detected in 2004 only.}
	\label{fig:6196}
\end{figure}

\begin{figure}
	\centering
		\includegraphics[width=\columnwidth]{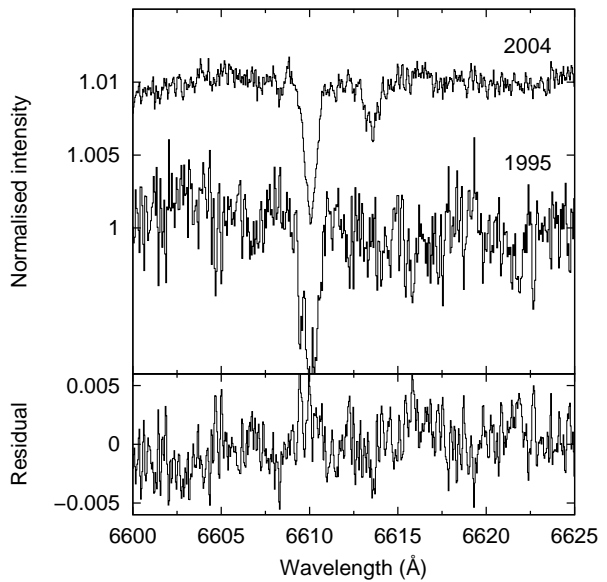}
	\caption{As in figure~\ref{fig:5780}, but for the $\lambda$6613 DIB. The line at 6610\,\AA\ is due to stellar N\,\textsc{ii}, whilst the DIB is detected in 2004 only.}
	\label{fig:6613}
\end{figure}

A search was performed for those DIBs in the observed wavelength regions with potentially detectable equivalent widths, by reference to the survey of diffuse interstellar bands by \citet{Hobbs2008}. Five DIBs ($\lambda\lambda$5780, 6196, 6203, 6283 and 6613) were detected towards \kvel\ in the 2004 data; only $\lambda5780$ was detected in the lower signal-to-noise 1995 data (see figures~\ref{fig:5780}--\ref{fig:6613}). As far as the authors are aware, these are the first reported detections of DIBs towards this star, so there are no other epochs for comparison.

The measured DIB equivalent widths are given in table~\ref{tab:dibs}, along with \nai~D$_1$ which was present in both observations, and the \ki\ 7698 line which was within the wavelength coverage in 2004 only. Uncertainties and upper limits were calculated as $\sigma_c\Delta\lambda$, where $\sigma_c$ is the continuum noise level and $\Delta\lambda$ the DIB FWHM. This simple method is fairly conservative and probably over-estimates the uncertainties, particularly for the broadest bands. The $\lambda6283$ DIB overlaps the telluric O$_2$ a-band, which could only be adequately removed from the 2004 data. Blends with stellar lines were determined by reference to the stars $\upsilon$~Sco and $\alpha$~Pav (both B2\,IV), which were observed on the same night as the 2004 data. For $\lambda6283$, a narrow stellar line at 6287\,\AA\ was removed via a Gaussian fit; there are possible weak stellar blends with $\lambda5780$ and $\lambda5797$. The displacement in the continuum adjacent to 5780 is likely to have been introduced as a result of weak stellar features in the telluric standards.

As a comparison between the two epochs, the spectra around $\lambda5780$ (the only DIB detected at both epochs, and including the region where $\lambda5797$ would be expected) were subtracted and the residuals evaluated. This procedure is shown in figure~\ref{fig:5780}; within the noise there has apparently been no change in $\lambda5780$ between January 1995 and June 2004. From the residuals and the uncertainties in the measured equivalent widths, we determine an upper limit on the change in $\lambda5780$ between the two epochs of $\la$\,40\%.

\section{Discussion}
\label{sec:discuss}

\subsection{Comparison of observations at different epochs}
\label{sec:compare}

\subsubsection{Equivalent widths}
\begin{figure}
	\centering
		\includegraphics[width=\columnwidth]{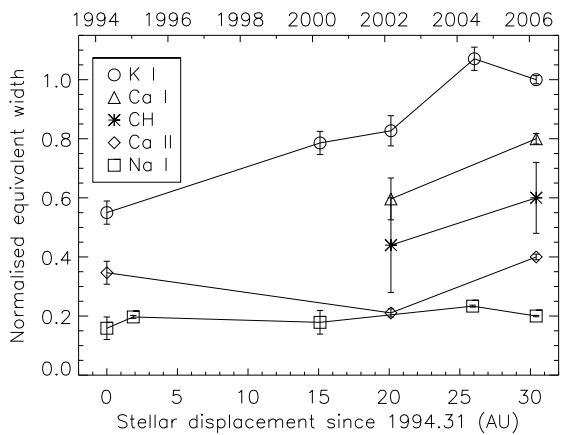}
	\caption{Comparison of observed equivalent widths of interstellar absorbers towards \kvel\ at different epochs. Equivalent widths of the interstellar absorption in 1994, 2000 \citep{Crawford2000}, 2002 \citep{Crawford2002}, 1995, 2004 and 2006 (this work) are plotted against stellar displacement on the bottom axis, and epoch in years on the top axis. Data have been normalised with respect to the 2006 values and vertical offsets of 0.2 between species applied for display; lines have been added to guide the eye.}
	\label{fig:ew}
\end{figure}

A model-independent comparison of the measured equivalent widths of each atomic and diatomic line from all available epochs since 1994 is presented in figure~\ref{fig:ew}. For comparison with the earlier measurements from \citet{Dunkin1999}, \citet{Crawford2000} and \citet{Crawford2002}, we have consistently compared equivalent width measurements made over the same velocity range.

For \ki, the combined equivalent width of only the narrower $\sim9$\kms\ components has been plotted, to be consistent with the earlier data; this was measured as $5.09 \pm 0.07$\,m\AA\ in the 2006 data by first dividing out the model fit to the broad, higher velocity \ki\ components. These broad components are unavoidably blended with the narrow components in the lower-resolution 2004 UCLES data, leading to that datum appearing too high in figure~\ref{fig:ew}. To check this, the 2006 \ki\ observation (including the broad components) was degraded to the same resolution as the 2004 observation, and the equivalent width remeasured. The resulting value was 5.9\,m\AA, which is somewhat lower than the total 7.49\,m\AA\ measured from the UHRF data (table~\ref{tab:uhrfobs}) but still higher than the 5.09\,m\AA\ of the narrow components alone. The equivalent width measured at the degraded resolution is thus similar to that measured in the 2004 UCLES data (see table~\ref{tab:dibs}); so the apparently inflated 2004 datum could be an effect of the lower resolution.

There are significant increases in equivalent width for the \ki\ and \cai\ lines. $W(\text{\ki})$ increased by $82\pm7\%$ between 1994 and 2006, whilst $W(\text{\cai})$ increased by $26\pm9\%$ between 2002 and 2006. The equivalent width of \nai\ barely changed over the same time periods ($4\pm4\%$ between 1994 and 2006, and $2\pm4\%$ between 2002 and 2006), although this may be due to the near-saturation of the \nai\ absorption, which lies on the flat part of the curve of growth. The measured \nai\ equivalent width also includes the broad, higher velocity components which are not suspected to be variable, given the lack of any significant differences between the models of the 1994, 2000 and 2006 data (see section~\ref{sec:nai}). \caii\ shows no significant increase ($6\pm4$\%) in equivalent width between 1994 and 2006, at least when matching velocity components are compared; the 2006 measurement plotted here ($20.6\pm0.2$\,m\AA) excludes the $\sim 0$\kms\ component, but the (apparently low) 2002 datum is almost certainly compromised by interference from stellar \caii\ K, which was well separated from the interstellar \caii\ K line in the 1994 and 2006 data. The increase in CH equivalent width is not significant at $19\pm22\%$ between 2002 and 2006, but nor are the data inconsistent with CH matching the increases seen in \ki\ and \cai.

\subsubsection{Column densities}
\label{sec:compare_N}
\begin{figure}
	\centering
		\includegraphics[width=\columnwidth]{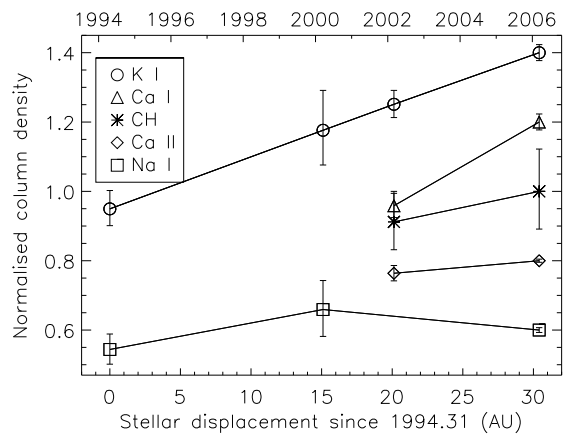}
	\caption{As figure~\ref{fig:ew}, but for the combined column density of the interstellar absorption at $\sim$\,9\kms. Only the UHRF data from 1994, 2000, 2002 and 2006 are plotted; error bars for 1994 and 2000 are probably overestimates by factors of 5--10 (see section~\ref{sec:compare_N}).}
	\label{fig:N}
\end{figure}

The problems caused by the near-saturation of the \nai\ line and the inclusion of the broad $\sim$\,12\kms\ component may be avoided by instead comparing the modelled column density at each epoch, and restricting comparison to the $\sim$\,9\kms\ component(s) only. This is impractical for the Mt. Stromlo and UCLES \ki\ data, in which the narrow and broad components are not resolved, so they are not considered further. The individual (A and B) components in the UHRF data cannot always be directly compared with the previously published column densities because the absorption has been modelled with different numbers of components (e.g. for \cai\ and \caii). However, all the remaining observations were made with the same instrument at essentially the same spectral resolution (850,000--900,000), and the total column densities (A\,+\,B) in the 9\kms\ components are robustly determined.

In figure~\ref{fig:N}, the components reported in \citet{Crawford2000,Crawford2002} and in this work are compared by summing the column densities of the one or two components which make up the absorption at $\sim$\,9\kms. Uncertainties in the total column density obtained simply by summing individual components in quadrature are overestimates, because in overlapping lines the column densities of the two components are not independent quantities. The 2002 observations are not affected, because they only utilised one component. For the 2006 observations, the uncertainties were obtained by interrogating the Monte-Carlo error simulations, determining the sum of column densities at each iteration, and taking the 68\% range. This process revealed a strong anticorrelation between the components, and the total column densities were found to be $\log{N}_{\text{AB}}(\text{\cai})=9.62\pm0.01$, $\log{N}_{\text{AB}}(\text{\caii})=11.316\pm0.002$, $\log{N}_{\text{AB}}(\text{\ki})=10.47\pm0.01$ and $\log{N}_{\text{AB}}(\text{\nai})=11.985\pm0.003$. For the 1994 and 2000 observations, only summing in quadrature is possible; by comparison with the 2006 observations these error bars are likely to be overestimates by factors of 5--10.

The differing treatments of the higher velocity absorption may still affect this comparison. For example, the broad \ki\ components partially overlap the $\sim9$\kms\ absorption, but were not included in the model of \citet{Crawford2002}; this has the effect of slightly enhancing the apparent \ki\ column density in 2002.

Figure~\ref{fig:N} shows that the increase in \ki\ column density first noted by \citet{Crawford2000} has continued until 2006: $N(\text{\ki})$ has increased by $82^{+10}_{-9}\%$ between 1994 and 2006. For the first time, \cai\ is found to have increased -- by $32\pm5\%$ between 2002 and 2006 (a slightly higher rate than \ki\ over the same period). \caii\ and CH show no statistically significant changes between 2002 and 2006, and nor does \nai\ over the period 1994--2006. \cai\ and \caii\ are therefore seen to vary at different rates to each other, indicating a change in the ionisation conditions.

\subsection{Line widths and temperature}
\label{sec:temp}
Although one of the aims of the UHRF observations was to measure the kinetic temperature from the rotational populations of \ctwo, the lack of a detection of this molecule necessitates the use of another method. In addition to the instrumental resolution and the natural line width, the two mechanisms which act to broaden interstellar lines are thermal Doppler broadening and turbulent bulk motions of the gas. These are related to the Doppler broadening parameter $b$ through:
\begin{equation}
	b=\sqrt{\frac{2k_BT_k}{m}+v_t^2}
	\label{eq:b}
\end{equation}
where $k_B$ is the Boltzmann constant, $T_k$ is the kinetic temperature, $m$ is the atomic (or molecular) mass of the species and $v_t$ is the RMS turbulent velocity. Taking the limiting case of zero turbulent velocity, rigorous upper limits on the kinetic temperature $T_k^{\rm{ul}}$ may be obtained for each species, and are given in table~\ref{tab:uhrfmodels} when $T_k^{\rm{ul}}<$\,2,000\,K.

In principle it should be possible to determine $T_k$ and $v_t$ separately by solving equation~\ref{eq:b} for species with differing values of $m$. However, there are no systematic differences in $b$ values between species in component B, over a range $m=23$ (\nai) to $m=40$ (\cai\ and \caii). In component A, statistically significant differences in $b$ are found between species, but with \ki\ the narrowest and \cai\ the broadest. This is difficult to reconcile with equation~\ref{eq:b}, which implies that the least massive species (Na) should have the highest $b$ value. The most likely explanation for these discrepancies is that the species are not entirely physically coincident and reside within different parts of the cloud, perhaps in regions with differing $v_t$. If $v_t$ makes a significant contribution to the $b$ values, then the kinetic temperature could be substantially lower than the derived upper limits.

\subsection{CH distribution}
\label{sec:CH_sim}
\begin{figure}
	\centering
		\includegraphics[width=\columnwidth]{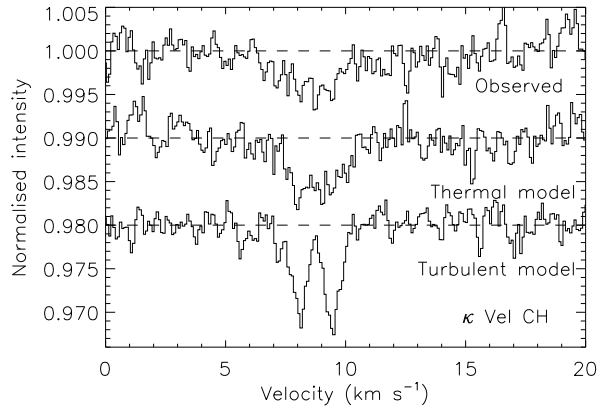}
	\caption[Observed and simulated CH line profiles]{Observed and simulated CH line profiles. Upper histogram is the observed CH profile as in figure~\ref{fig:CH}, middle histogram is the `thermal' model, and the lower histogram is the `turbulent' model, assuming CH shares the same distribution as \ki\ (see section~\ref{sec:CH_sim} for details of the models). The spectra have been vertically offset by 0.01 for display; the dashed lines illustrate the continuum levels. The observed line profile is significantly broader than the turbulent model, and the $\Lambda$-doublet is not visible in the observed spectrum.}
	\label{fig:CH_sim}
\end{figure}

Visual inspection of the CH spectrum shows no sign of the expected $\Lambda$-doublet line profile, which causes the relatively high value of $b$ in the single-component \textsc{vapid} model. In order to determine if CH has the same line profile as the atomic species but the $\Lambda$-doublet is being hidden by the noise, simulated CH line profiles were constructed. Two components were assumed, with velocity separation and column density ratio equal to that observed in \ki, which is known to be distributed similarly to CH in diffuse clouds \citep[e.g.][]{Welty2001,Pan2005}, and a total column density equal to the CH model. Two limiting cases for $b$ were considered: $v_t$ dominant (the `turbulent' model, $T_k=0$), and $T_k$ dominant (the `thermal' model, $v_t=0$). For the turbulent model we assume $b$ values equal to those of \ki\ (table~\ref{tab:uhrfmodels}). Because CH ($m=13$) is substantially less massive than K ($m=39$), the thermal model adopts $b$ values of 1.06 and 0.60\kms\ i.e. a factor of $\sqrt{3}$ ($=\sqrt{39/13}$) higher than the turbulent model. These simulated line profiles were then degraded to the obtained noise level using clean sections of the observed continuum near the CH line, which avoids the need to make assumptions about the noise statistics.

The simulated spectra are compared with the observation in figure~\ref{fig:CH_sim}. The $\Lambda$-doublet is clearly visible in the turbulent model, but barely discernible in the thermal model. A simulation in which only component A is `turbulent' is not very different from the combined thermal model illustrated: it is the degree of turbulent broadening in the narrower component B which is primarily constrained by this analysis. It is therefore plausible that the CH shares a similar distribution to \ki\ but the line profile is being masked by noise. If so, the turbulent contribution to the CH line profile is low, at least in component B. This suggests that the temperature of component B cannot be much lower than the upper limit derived in section~\ref{sec:temp}.

We also constructed simulated spectra based on the \cai\ distribution, which is more heavily concentrated in component~B than \ki. In those simulations the $\Lambda$-doublet is clearly visible even in the thermal model, which is inconsistent with the observations. This supports the assumption made above that CH is distributed more similarly to \ki\ than \cai.

\subsection[Ca I / Ca II ratio and electron density]{Ca\,{\sevensize I} / Ca\,{\sevensize II} ratio and electron density}
\label{sec:ca_ratio}
At the low densities prevalent in the diffuse ISM, the balance between the various ionisation states of a single atom is set by the competition between photoionisation which promotes the atom to a higher ionisation state, and recombination with an electron which demotes it to a lower one. For the neutral and first ionised state in equilibrium, this may be expressed as:
\begin{equation}
	\Gamma(\text{X\,{\sc i}})\,n(\text{X\,{\sc i}}) = \alpha_r(\text{X\,{\sc i}},T)\,n(\text{X\,{\sc ii}})\,n_e
	\label{eq:ion_balance}
\end{equation}
where $\Gamma$ is the photoionisation rate, $\alpha_r$ the radiative recombination rate coefficient, $n_e$ the electron number density and $n(\text{X})$ the number density of species X. As observations provide the column density $N$ and not the number density $n$, it is necessary to assume a constant density along the line of sight, allowing $n$ to be replaced by $N$. Under this assumption, and for the specific case of calcium, equation~\ref{eq:ion_balance} may be rewritten as:
\begin{equation}
	n_e = \frac{\Gamma(\text{\cai})}{\alpha_r(\text{\cai},T)}\, \frac{N(\text{\cai})}{N(\text{\caii})}.
	\label{eq:ca_ratio}
\end{equation}

Calcium is an unusual case because both its first and second ionisation energies (6.11, 11.87\,eV respectively) are lower than the Rydberg energy (13.6\,eV), so in diffuse clouds ionisation to \caiii\ is possible. There is then a second ionisation equilibrium between \caii\ and \caiii, though this does not invalidate the electron densities derived from equation~\ref{eq:ca_ratio}. \citet{Welty2003} calculated the relative populations of \cai, \caii\ and \caiii\ and found that the dominant ionisation state is \caii\ in any situation where \cai\ is abundant enough to be detected.

Equation~\ref{eq:ca_ratio} is an approximation which assumes that no other processes cause significant conversion between ionisation stages. Plausible processes include cosmic-ray ionisation and charge exchange with dust grains or PAHs \citep[e.g.][]{Liszt2003}. \citet{Welty2003} have discussed this issue in the context of calcium ionisation ratios, and concluded that equation~\ref{eq:ca_ratio} remains valid if the fractional ionisation ${n_e}/{n_H}\la2\times10^{-4}$, which is the level provided by complete ionisation of gas-phase \ci\ to \cii\ (assuming standard depletions, see also section~\ref{sec:density}). Those authors also conclude that the assumption of photoionisation is most valid when $n_e$ is high. Since the observations towards \kvel\ indicate a cloud which is much cooler and/or more dense than usual for the diffuse ISM, and as the low extinction present in the cloud\footnote{$\ebv=0.10$, so $A_V=0.16$ halfway through the interstellar material along the line-of sight, assuming $R_V=3.1$. The transverse extinction is unknown but probably even lower.} favours photoionisation, the assumptions leading to equation~\ref{eq:ca_ratio} are expected to be valid in this case.

The electron density may then be determined from the column densities, given the value $\Gamma/\alpha_r=66$ for \cai\ given by \citet[their table 3]{Welty2003}\footnote{This value assumes $T=100$\,K. For calcium, the dependence of $\alpha_r$ on $T$ is $\alpha\propto(T/100\text{\,K})^{-0.683}$ \citep{Pequignot1986}. $\Gamma$ is independent of $T$.}. The electron densities are then found to be $n_{e,\text{A}}=0.98\pm0.12$\pcc and $n_{e,\text{B}}=3.14\pm0.77$\pcc\ for components A and B respectively. These are very high values for diffuse clouds. For comparison, \citet{Crawford2002} found $n_e=0.97$\pcc\ towards \kvel\ using a single-component model.

Some observers have found that electron densities derived from calcium are sometimes higher than those derived from other species along the same lines of sight. \citet{Welty1999} used ionisation ratios of nine different elements to determine $n_e$ towards 23~Ori, with calcium resulting in the highest estimate, whilst \citet{Sonnentrucker2003} found that calcium gave the highest $n_e$ (out of four measured ionisation ratios) towards HD~185418. This effect has only been reported towards a handful of stars, and the cause is unknown (see \citealt{Welty2003} for a discussion). There is some circumstantial evidence to support the electron density derived from calcium: the changes seen in \ki\ are about the same rate as those seen in \cai\ (section~\ref{sec:compare}), suggesting that both are responding to the same change in (observed) electron density; and the total number densities derived independently from the electron density and the chemical model (sections~\ref{sec:density} and \ref{sec:chemical_model}) are in agreement with each other.

\subsection[Ca I / K I ratio and depletions]{Ca\,{\sevensize I} / K\,{\sevensize I} ratio and depletions}
\label{sec:ca/k}
Whilst the \cai/\caii\ ratio provides a measure of the electron density (under the assumption of constant temperature), some line ratios are sensitive to other properties of the interstellar gas. Because the depletion of potassium into dust grains is thought to be fairly constant, the \cai/\ki\ ratio may be used to estimate the depletion of calcium, which is typically very high at the low temperatures in the diffuse ISM due to its high condensation temperature ($T_C=1,659$\,K, \citealt{Lodders2003}). \citet{Welty2003} found that $N(\text{\cai})$ and $N(\text{\ki})$ are correlated, with the slope of that correlation arising from the difference in depletion behaviours. Our measured ratios fall very close to the mean relation determined by \citet{Welty2003}, suggesting that there is no unusual enhancement of \cai\ along this line of sight. Adopting the same assumptions as for the calcium ionisation balance, and assuming that \caii\ is dominant, the relative populations of \cai\ and \ki\ are given by:
\begin{equation}
\frac{N(\text{\cai})}{N(\text{\ki})}=\frac{(\Gamma/\alpha_r)_{\text{\cai}}}{(\Gamma/\alpha_r)_{\text{\ki}}} \frac{\delta(\text{Ca}) a_{\text{Ca}}}{\delta(\text{K}) a_{\text{K}}}
\end{equation}
where $\delta(X)$ is the gas-phase fraction and $a_X$ the cosmic abundance of species $X$. Taking the standard abundances, $\delta(\text{K})=0.2$, and photoionisation and recombination rates given by \citet{Welty2003} this becomes
\begin{equation}
	\frac{N(\text{\cai})}{N(\text{\ki})}\approx12.4\delta(\text{Ca}).
	\label{eq:ca_k_ratio}
\end{equation}
This gives estimates of the calcium gas-phase fraction in components A and B of $\delta(\text{Ca})_\text{A}=(1.51\pm0.23)\times10^{-2}$ and $\delta(\text{Ca})_\text{B}=(7.42\pm1.45)\times10^{-3}$, which may be equivalently expressed as $D(\text{Ca})_\text{A}=-1.82\pm0.13$ and $D(\text{Ca})_\text{B}=-2.13\pm0.17$, where $D$ are the logarithmic depletions. \citet{Cardelli1991} investigated the depletion of calcium in diffuse clouds, and concluded that it is very sensitive to the local density, increasing in denser regions. This is in accordance with our result of a higher calcium depletion in component B (which is found to be approximately a factor of two denser than component A in section~\ref{sec:density}).

\subsection{Molecular hydrogen fraction}
\label{sec:h2}

The \cai/\ki\ ratio is also correlated with the fraction of hydrogen in the form of \htwo, though the relation is not particularly tight \citep{Welty2003}. Previous observations have only resulted in upper limits for the \hi\ and \htwo\ column densities towards \kvel\ (\citealt{Jenkins2009}, whose results were based on the observations of \citealt{Bohlin1983} corrected for stellar contamination). The measured ratios are $\log(N(\text{\cai})/N(\text{\ki}))_\text{A}=-0.729\pm0.035$ and $\log(N(\text{\cai})/N(\text{\ki}))_\text{B}=-1.036\pm0.052$. These correspond to upper limits for the fraction of hydrogen in the form of \htwo\ of $f(\text{\htwo})\la10^{-2}$ in both components \citep[their figure 4]{Welty2003}. These are very low molecular hydrogen fractions for the diffuse ISM, especially for regions with enhanced density.

A more direct estimate of the molecular hydrogen fraction could in principle be obtained from the standard relations between \ebv\ and \hi, and between $N(\text{CH})$ and \htwo. Adopting $\ebv=0.10$ \citep{Cha2000} with an uncertainty\footnote{The uncertainty on \ebv\ is taken to be equivalent to a misclassification by one spectral subtype.} of 0.02, equation 7 of \citet{Burstein1978} predicts $\log N(\text{\hi})=20.9\pm0.2$. Taking $\log(N(\text{CH})/N(\text{\htwo}))=-7.46\pm0.21$ from \citet{Sheffer2008}\footnote{Note that this requires extrapolating the  \citet{Sheffer2008} relation to lower column densities than they actually observed.}, we obtain an estimate of $\log N(\text{\htwo})=18.93\pm0.22$. This implies a molecular fraction of $f(\text{\htwo})\equiv\frac{2N(\text{\htwo})}{N(\text{\hi})+2N(\text{\htwo})}=(2.1\pm1.6)\times10^{-2}$, which is in line with the previous estimate.

Both of these estimated column densities are more than one dex higher than the upper limits from the UV observations \citep{Jenkins2009}. We also note that \citet{Welty2001} presented empirical relations between \ki, \nai\ and total hydrogen column density for typical diffuse ISM lines of sight, but for \kvel\ these would again imply at least an order of magnitude more hydrogen than the observed upper limits. Apparently the standard relations are not valid for the line-of-sight towards \kvel. It is possible that the column densities have changed by this amount between the epochs of the UV observations (late 1970s, \citealt{Bohlin1983}) and the optical observations (the \ebv\ measurement is based on Tycho photometry so has an effective epoch of 1991.5, whilst the spectroscopic measurements are from 2006). Alternatively the ISM towards \kvel\ could be out of chemical equilibrium \citep{Bell2005} and thus these standard relations might not apply.

\subsection{Physical dimensions and density}
\label{sec:density}
The proper motion of \kvel\ is 15.5\,mas\,yr$^{-1}$ \citep{Perryman1997}, so at a distance of 165\,pc the transverse velocity is 2.56\,AU\,yr$^{-1}$ (12.1\kms). This gives a stellar displacement of 10.1\,AU between the 2002 and 2006 observations, and 29.7\,AU between 1994 and 2006 (shown on the axes of figures \ref{fig:ew} and \ref{fig:N}). The size of the binary orbit has been determined to be $a_1\sin i=0.48$\,AU, where $a_1$ is the semi-major axis of the orbit of the primary, and $i$ the orbital inclination \citep{Pourbaix2004}. Since neither the inclination nor position angle of the orbital ellipse have been determined, it is impossible to include the effect of this on the transverse displacement, but the error thus introduced is likely to be rather small.

The displacement of the line-of-sight at the distance of the star provides an upper limit on the displacement at the distance of the foreground absorbing cloud. \citet{Dunkin1999} have argued that the cloud is likely to be located close to the star, at least 150\,pc from the Sun, based on comparison with CO maps and optical absorption along other nearby lines-of-sight \citep[see also][]{Crawford1991}. This would place the cloud just beyond the edge of the Local Bubble \citep{Welsh2010}. Any transverse motion of the interstellar cloud would modify the displacement of the line-of-sight through the cloud by the vector sum of the motions of the star and cloud. It is impossible to determine the transverse velocity of the cloud, but reasonable assumptions suggest that the derived cloud size is accurate to within a factor of 2--3, in either direction.

If the electron density is dominated by electrons released through the photoionisation of atomic carbon (the assumption adopted by \citealt{Crawford2002}), the electron density allows an estimate of the total number density to be obtained. Assuming a carbon depletion of 60\% implies a gas-phase carbon abundance of $1.4\times10^{-4}n_\text{H}$ \citep{Sofia1997}, which leads to estimates of $n_\text{A}\ga7\times10^3$\pcc\ and $n_\text{B}\ga2\times10^4$\pcc. These are orders of magnitude higher than typical values for diffuse or translucent clouds \citep{Snow2006}. They are also higher than the densities inferred by some small-scale structure studies \citep[e.g.][]{Pan2001}, but those studies relied upon simple chemical models to determine the density. \citet{Bell2005} demonstrated that the chemistry can be time-dependent in small-scale structures, whilst the atomic species react much more rapidly to changing conditions.

\citet{Welty2007} combined optical and UV observations (including \ci\ and \cii) of the small-scale structure towards HD~219188. It was inferred that the changing column densities reflected enhanced recombination due to a raised electron density, the source of which was the ionisation of hydrogen. If this effect is also present in the line-of-sight towards \kvel, the densities derived above will be overestimates. Since hydrogen is thousands of times more abundant than carbon, even a small fractional ionisation of hydrogen could reduce the density estimate by an order of magnitude or more. An ionisation fraction of $\ga1\%$ would be required to reduce the density to levels typical of the diffuse ISM. A physical cause of the ionisation would also be required, such as intense cosmic ray bombardment or penetration of soft X-rays through the relatively low-extinction gas \citep{Tielens2005}, though the latter might plausibly arise from the hot gas in the Local Bubble \citep{Frisch2011}.

An estimate of the hydrogen column density may be derived from the observed metal column densities and depletion estimates. Taking $\log[\text{Ca}/\text{H}]+12=6.34$ from \citet{Asplund2009}, D(Ca) estimates from section~\ref{sec:ca/k} and the observed \cai\ + \caii\ column densities, we obtain $\log{}N_\text{A}($H$)\approx18.7$ and $\log{}N_\text{B}($H$)\approx18.3$, so $\log{N}_{\text{AB}}($H$)\approx18.9$. This is consistent with the observational upper limits of $\log{N}($\hi$) < 19.5$ and $\log{N}($\htwo$) < 17.7$ \citep{Jenkins2009}. Taking these column density estimates and assuming a density of $10^4$\pcc\ gives an estimated total path length of $\sim50$\,AU. This length estimate rests upon a series of assumptions, in particular the derived calcium depletions and neglecting \caiii, so is probably only indicative to within an order of magnitude.


\subsection[C_2 column density]{\ctwo\ column density}
\label{sec:c2_col}
The relative populations of the low-$J$ rotational states of \ctwo\ depend upon the local density and kinetic temperature. In order to determine an upper limit on the total \ctwo\ column density it is necessary to combine the limits on the equivalent width for lines with differing $J$ derived in section~\ref{sec:c2_obs} with the density and temperature determinations from sections~\ref{sec:temp} and \ref{sec:density}. Since \ctwo\ is expected to preferentially form in denser regions, we adopt $n=2\times10^4$\pcc\ and assume $T_k=100$\,K.

The relative populations of each $J$ level were calculated using B.~McCall's online calculator\footnote{\url{http://dib.uiuc.edu/c2}}, which uses the method of \citet{vanDishoeck1982}. The populations of each level were combined with the oscillator strengths from table~\ref{tab:uhrfatomicdata} and the calculated total population (i.e. integrating over all $J$) to determine the maximum total column density which is compatible with the equivalent width upper limits. The resulting $2\sigma$ upper limit is $N(\text{\ctwo})\la5.6\times10^{11}$\pcm.

Adopting a higher density or lower temperature would result in a somewhat tighter upper limit. At $T_k=100$\,K the calculation is rather insensitive to $n$: if $n$ changes by an order of magnitude the calculated upper limit differs by less than 5\%. Varying $T_k$ by 20\,K changes the upper limit by $\sim$\,10\%.

\subsection{Comparison with chemical models}
\label{sec:chemical_model}
The chemical models of \citet{Bell2005} made predictions for the column densities of CH, \ctwo, \cai\ and \caii\ (plus OH and CO) along a line of sight through an SSS filament. The physical parameters were based on those derived by \citet{Crawford2002} for \kvel, namely an extended structure of length 100\,AU along the line-of-sight and transverse depth of 10\,AU. Results were presented for a grid of cloud densities and times.

Because CH shows no evidence for more than one cloud component and \ctwo\ has only an upper limit over the total line-of-sight, for comparison with the chemical model column densities were initially summed over both components A and B. The total column densities adopted in this case\footnote{Although components A and B were not observed in CH and \ctwo, we assume that these molecules are confined to those cloud components.} were $N_{\text{AB}}(\text{CH})=2.95\times10^{11}$\pcm, $N_{\text{AB}}(\text{\ctwo})\la5.6\times10^{11}$\pcm\ ($2\sigma$), $N_{\text{AB}}(\text{\cai})=4.17\times10^9$\pcm\ and $N_{\text{AB}}(\text{\caii})=2.07\times10^{11}$\pcm\ (cf. section~\ref{sec:compare}). With the exception of \caii, these column densities are fully consistent with the \citeauthor{Bell2005} results for $n=5\times10^4$\pcc, $t=10$--50\,yr. Models with $n=10^5$\pcc\ are excluded by the \ctwo\ upper limit, whilst those with lower density are excluded by the \cai\ column density. The observed \caii\ column density is moderately higher than the model, by a factor of $\sim$\,4, but this is within the uncertainties of the models.

In an attempt to include material only in the densest regions, we have also compared the \citet{Bell2005} models to component B alone -- this component was suspected to be variable by \citet{Crawford2000}. It then becomes necessary to assume that all of the CH arises within component B (but see the discussion in section~\ref{sec:ch}), and the high column density of \cai\ in component A cannot be explained. Bearing in mind these caveats, the CH and \ctwo\ column densities remain as above whilst the calcium column densities are $N_\text{B}(\text{\cai})=1.58\times10^9$\pcm\ and $N_\text{B}(\text{\caii})=3.32\times10^{10}$\pcm. All the observed column densities are then consistent with the \citeauthor{Bell2005} model for $n=5\times10^4$\pcc, $t=10$\,yr.

The chemical models thus provide confirmation of the high number density in the cloud towards \kvel, and are most consistent with densities only slightly higher than the lower limits derived in section~\ref{sec:density}. The predicted \ctwo\ column density in the most consistent model is roughly a factor of three times lower than our observed upper limit. The models determined the local temperature through a self-consistent treatment of heating and cooling processes. The range of temperatures in the models was $\sim$\,30--115\,K, which are consistent with the upper limits derived in section~\ref{sec:temp}. It is difficult to interpret the time dependent aspect of the models, because the structure cannot have formed instantaneously. Nevertheless, the models predict that for $n=5\times10^4$\pcc\ CH and \caii\ would be essentially constant over time whilst \cai\ would increase, as observed (cf. figure~\ref{fig:N}).

\citeauthor{Bell2005} set $n(\text{\htwo})/n(\text{\hi})=0.4$, which is more than an order of magnitude higher than the molecular hydrogen fraction inferred in section~\ref{sec:h2}. Low molecular hydrogen fractions would act to reduce the column densities of the molecular species because the formation pathways all include reactions with \htwo. The estimates in section~\ref{sec:h2} might therefore be misleading, as the correlations may not hold in such high-density regions. Alternatively, the local density and/or \htwo\ fraction could be highly variable along the line-of-sight. A large number of such variations in local density (and to an extent temperature) have been proposed by \citet{Cecchi2009} as a solution to several problems in diffuse cloud chemistry. If this were the case, it would be possible for CH to be formed in regions of relatively high \htwo\ density, but for the atomic lines to trace less-dense material elsewhere, therefore reducing the total molecular fraction along the line of sight. In the absence of direct observations, this contradiction in $f(\text{\htwo})$ cannot currently be resolved.

\subsection{Diffuse interstellar bands}
The measured DIB equivalent widths are far lower than expected given the reddening, and only a few DIBs have been detected. Based on 133 lines of sight, \citet{Friedman2011} found a mean $W(\lambda5780)/\ebv$ value of $505$\,m\AA\,mag$^{-1}$. However, towards \kvel\ the reddening is $\ebv=0.10\pm0.02$ but $W(\lambda5780)$ is $10\pm1$\,m\AA, i.e. five times lower than would be typical for the Milky Way ISM. The relative weakness of $\lambda5780$ indicates a lower abundance of its carrier in the dense material towards \kvel\ than in the ambient ISM, which may indicate a density dependence of the (unknown) carrier formation mechanism.

Those DIBs which have been detected towards \kvel\ are members of \citet{Krelowski1987} group~2, whilst the DIBs of group 3 are absent. The equivalent width ratio of $\lambda5780$ (a group 2 DIB) to $\lambda5797$ (group 3) is at least $W(\lambda5780)/W(\lambda5797)>22$ ($1\sigma$), which is amongst the highest known: typical values of this ratio are 1--4. As far as the authors are aware, the highest ratios previously reported in the literature are $W(\lambda5780)/W(\lambda5797)>9$ ($1\sigma$) towards the star $\epsilon$\,Cassiopeiae by \citet{Friedman2011}, and $W(\lambda5780)/W(\lambda5797)=8.2$ (no uncertainty given) towards HD~20336 by \citet{Galazutdinov1998}\footnote{\citet{Porceddu1992} report five sightlines with high $\lambda5780$\slash{}$\lambda5797$ ratios, and \citet{Galazutdinov1998} three more, but neither set of authors report the value of the ratio nor provide numerical upper limits on $W(\lambda5797)$.}. \citet{Krelowski1999} found a correlation between $W(\lambda5797)/W(\lambda5780)$ and $W(\text{CH})/\ebv$. Although the value of $W(\text{CH})/\ebv=2.5\pm0.6$\,m\AA\,mag$^{-1}$ measured towards \kvel\ is much lower than any measured by \citeauthor{Krelowski1999}, the upper limit on $W(\lambda5797)/W(\lambda5780)$ does appear to fall on or near this relation.

The observed $\lambda5780/\lambda5797$ ratio corresponds to a cloud of extreme $\sigma$-type \citep{Krelowski1992}, which is generally interpreted as a low density, high UV radiation field environment. The additional DIBs and diatomic molecules present in $\zeta$-type clouds (generally interpreted as cooler, denser and less UV-irradiated than $\sigma$-type) are extremely weak or absent. However, the analysis above indicates that the interstellar cloud(s) toward(s) \kvel\ is/are relatively dense but unshielded from UV radiation. The weakness of $\lambda5780$ with respect to \ebv\ and the extreme $\sigma$-type line-of-sight suggests that unusual conditions are present in the cloud towards \kvel. \citet{Cami1997} found similar properties (weak DIBs per unit \ebv, extreme $\lambda5780$ to $\lambda5797$ ratio, weak or absent diatomics) towards targets in Orion, where the conditions are also fairly dense but highly UV irradiated. \citet{Snow1995} found systematically weak DIBs in reflection nebulae, where conditions are even more extreme in terms of high UV field and high electron density.

It has been suggested that the carrier of the $\lambda5797$ DIB is a neutral species which is easily ionised, whilst the $\lambda5780$ carrier does not easily change ionisation state due to being either an already ionised species or a neutral which is highly resistant to ionisation \citep[e.g.][]{Cami1997,Krelowski1998}\footnote{\citet{Snow1995} argued against already ionised species, instead favouring neutral molecules.}. Because the $\lambda5780$ intensity over a large sample correlates very well with $N(\text{\hi})$ it has also been suggested that the carrier has an ionisation energy similar to that of hydrogen \citep{Friedman2011}. Whilst the hypothesis on the nature of the $\lambda5797$ carrier is in accordance with the widely observed enhancement of $\lambda5797$ in less UV-irradiated environments, it appears to be at odds with the lack of $\lambda5797$ detection towards \kvel\ given the high electron densities derived in section~\ref{sec:ca_ratio}.

High electron densities should lead to enhanced recombination of molecular cations to form neutral molecules, as well as the observed neutral atomic species. But this would result in relatively strong $\lambda5797$, opposite to what is observed towards \kvel. Electronic recombination with a molecular cation does not necessarily produce the neutral through radiative association because the molecule may lose energy through fragmentation instead of emitting a photon. But if such fragmentation pathways are significant for the carrier of $\lambda5797$ this should result in \emph{destruction} of the carrier in  environments with high electron density; this is incompatable with the hypothesised enhancement of $\lambda5797$ in dense environments due to enhanced recombination of a cation.

It appears that the carrier of the $\lambda5797$ DIB (and by extension, the other DIBs in \citeauthor{Krelowski1987} group 3) is more sensitive to UV irradiation than density or degree of ionisation. The observations of \kvel\ may indicate that the carrier of $\lambda5797$ is dissociated by UV radiation more rapidly than the carrier of $\lambda5780$.

Although the \ki\ line was not observed in 1995, by interpolating figure~\ref{fig:ew} it is inferred that the equivalent width of the \ki\ line increased by $\approx80\%$ between the 1995 and 2004 observations; the limit on the change in $\lambda5780$ DIB is $\la40\%$. If the carrier of $\lambda5780$ were to have increased in proportion with the \ki\ line, a change would have been seen between 1995 and 2004, albeit with low statistical significance. \citet{Cordiner2006Fara} and Cordiner et al. (in prep.) found that DIB carriers (including $\lambda5780$) trace the same SSS as atomic lines such as \ki. The fairly loose upper limit on the variation of $\lambda5780$ towards \kvel\ is insufficient to draw firm conclusions on this point.

\section{Conclusions}
\label{sec:conclusion}
New ultra-high resolution observations of interstellar absorption towards \kvel\ by the species \ki, \nai, \cai, \caii\ and CH have been obtained. A search for \ctwo\ was undertaken, but no lines were detected. Best-fitting model line profiles were computed and two main narrow ($b<0.8$\kms) absorption components were identified. Comparisons with observations taken at earlier epochs show increases in interstellar \ki\ and \cai\ column densities over a period of several years. Since the previous observations in 2002, the transverse velocity of the star has shifted the line-of-sight through the interstellar cloud by $\sim$\,10\,AU.

From a consideration of the Doppler $b$ parameter in each component, a rigorous upper limit was placed on the kinetic temperature of each component. Combining limits from each species, the temperatures of components A and B must be $T_{k,\text{A}}<671^{+18}_{-17}$ and $T_{k,\text{B}}<114^{+15}_{-14}$\,K. However, the temperatures may be lower if the turbulent velocity makes a significant contribution to the $b$ values. The \cai/\ki\ ratio was used to derive logarithmic depletions of $D(\text{Ca})_\text{A}=-1.82$ and $D(\text{Ca})_\text{B}=-2.13$.

The \cai/\caii\ ratio was used to derive estimates of the electron density in the two components. By assuming that electrons originate from photoionisation of neutral atom\-ic carbon and assuming a value of 60\% for the carbon depletion, lower limits on the density of $n_\text{A}\ga7\times10^3$\pcc\ and $n_\text{B}\ga2\times10^4$\pcc\ were derived. These values are far higher than typical densities in the diffuse ISM. Comparison of the observations with the chemical model of \citet{Bell2005} confirms the requirement for a high number density, with $n=5\times10^4$\pcc\ in the best-fitting model.

The first detections of diffuse interstellar bands towards \kvel\ were made at two epochs, but only a loose upper limit of $\la40\%$ could be placed on the variation of DIB absorption between the two epochs. The DIBs detected are weak given the measured \ebv, and an unusually high $\lambda5780/\lambda5797$ ratio of $>22$ was found, amongst the highest known. This suggests that the carrier of $\lambda5797$ is more sensitive to UV radiation than local density. The behaviour of DIBs towards \kvel\ appears to be similar to that observed in Orion by \citet{Cami1997}. The weakness of the DIBs (per unit \ebv) may be related to the high density of the material towards \kvel.

Continued monitoring of interstellar absorption lines towards \kvel\ would be of interest, in order to determine the size of the structure from the time taken for the column density to peak and subsequently decline. It would be beneficial to observe the \ci\ fine structure lines in the ultraviolet to allow derivation of the local density and kinetic temperature. Although this approach leads to some degeneracy in these parameters, this could be lifted via the existing constraints on temperature and density, though (nearly) simultaneous optical observations would be an advantage. A direct measurement of the \htwo\ fraction would also be beneficial.

\section*{Acknowledgments}
The authors thank PATT for the award of UHRF time on the AAT and for T\&S. SJF and AMS thank Stuart Ryder and the AAO technical staff for their characteristically excellent support, and Julian Russell for his assistance with the AAT observations. Ian Crawford and Dan Welty provided helpful comments on early drafts of this paper, whilst Ian Howarth provided assistance with the \textsc{vapid} software. KTS acknowledges financial support from EPSRC, and MAC visitor funding from STFC.

\bibliographystyle{mn2e}
\bibliography{kVel_UHRF_v2}

\bsp

\label{lastpage}

\end{document}